\documentclass[%
superscriptaddress,
nofootinbib,
amsmath,amssymb,
aps,
]{revtex4-2}
\usepackage[colorlinks=true,linkcolor=blue,citecolor=blue,urlcolor=
blue]{hyperref}

\usepackage[utf8]{inputenc}
\usepackage{graphicx}
\usepackage{dcolumn}
\usepackage{bm}
\usepackage{amsmath}
\usepackage{tensor}
\usepackage[dvipsnames]{xcolor}
\usepackage{subcaption}
\usepackage{lipsum}
\usepackage{mwe}

\begin{document}
	
	\preprint{APS/123-QED}
	
	\title{Charged participants and their electromagnetic fields in an expanding fluid}
	
	\author{Ashutosh Dash}%
	\affiliation{Institut f\"ur Theoretische Physik, 
		Johann Wolfgang Goethe--Universit\"at,
		Max-von-Laue-Str.\ 1, D--60438 Frankfurt am Main, Germany}
	\email{dash@itp.uni-frankfurt.de}
	\author{Ankit Kumar Panda}
	\email{ankitkumar.panda@niser.ac.in}
	\affiliation{School of Physical Sciences, National Institute of Science Education and Research,
		An OCC of Homi Bhabha National Institute, Jatni-752050, India}%

	\date{\today}
	
	\begin{abstract}
		We investigate the space-time dependence of electromagnetic fields produced by charged participants in an expanding fluid. To address this problem, we need to solve the Maxwell's equations coupled to the hydrodynamics conservation equation, specifically the relativistic magnetohydrodynamics (RMHD) equations, since the charged participants move with the flow. To gain analytical insight, we approximate the problem by solving the equations in a fixed background Bjorken flow, onto which we solve Maxwell's equations. The dynamical electromagnetic fields interact with the fluid's kinematic quantities such as the shear tensor and the expansion scalar, leading to additional non-trivial coupling. We use mode decomposition of Green's function to solve the resulting non-linear coupled wave equations. We then use this function to calculate the electromagnetic field for two test cases: a point source and a transverse charge distribution. The results show that the resulting magnetic field vanishes at very early times, grows, and eventually falls at later times.
	\end{abstract}
	
	\maketitle
	\section{Introduction} 
	In heavy-ion collisions, intense transient electromagnetic fields are produced due to the motion of spectators, reaching magnitudes of approximately $\sim10^{18}-10^{19}$ G at RHIC-LHC energies~\cite{Kharzeev:2007jp,Skokov:2009qp,PhysRevC.83.039903,PhysRevC.83.054911,PhysRevC.85.044907,Tuchin:2013ie,McLerran:2013hla}. These intense electromagnetic fields may give rise to many novel phenomena, such as chiral magnetic effect (CME), chiral separation effect (CSE), etc~\cite{PhysRevD.83.085007,PhysRevD.78.074033,PhysRevLett.107.052303}. Detecting these phenomena in heavy-ion collisions is still an ongoing process. It is well-known that the bulk matter produced in heavy-ion collisions has low kinematic viscosity ($\eta/s$)~\cite{PhysRevLett.99.172301,PhysRevC.77.064901,PhysRevLett.94.111601}  and is well-described using the viscous relativistic hydrodynamics formulation~\cite{Romatschke:2017ejr,Luzum:2009sb,Romatschke:2007mq}. The relativistic generalization of first-order hydrodynamics in the Landau frame is acausal, so one needs to go to second-order in gradients of hydrodynamic fields to make the theory causal  \cite{ISRAEL1979341}. On a similar footing, for charged fluids, the generic framework is that of relativistic magnetohydrodynamics (RMHD) ~\cite{Hernandez:2017mch,Roy:2015kma}. As with uncharged fluids, the above framework has recently been extended to second-order in gradients of fluids and fields~\cite{Denicol:2018rbw,Denicol:2019iyh,Panda:2020zhr,Panda:2021pvq}. Other theoretical developments~\cite{Zhang:2022lje,Gezhagn:2021oav,Huang:2015oca,Karmakar:2022one,Panda:2022nsw} in this direction have also been made, along with numerical implementation for a comprehensive study of the bulk dynamics~\cite{Inghirami:2019mkc,Dash:2022xkz,Nakamura:2022wqr,Most:2021uck,Most:2021rhr,Dionysopoulou:2012zv,Huang:2022qdn,Das:2017qfi,Panda:2023akn}. For more detailed discussions on the developments in the field of RMHD one can follow~\cite{Hattori:2022hyo,Hernandez:2017mch}. 
	
	Nonetheless, all previous analyses focused on calculating the generation of electromagnetic fields from moving spectators. In Refs.~\cite{Kharzeev:2007jp,Gursoy:2018yai}, electromagnetic fields generated by participants are calculated from the geometric overlap region of the participants, using electromagnetic fields generated by the spectators. However, we believe that this method can be improved upon. Since charged participants constitute the bulk of the flow, the electromagnetic field generated by them requires thorough investigation, and that is the focus of our work. As previously mentioned, the generic framework for studying the dynamics of charged participants is RMHD, which can be challenging and often requires numerical simulation. Our approach here will be more modest, with a focus on analytical insights.  To this end, we approximate the problem by studying Maxwell's equations in a charged background fluid flow, which we take to be a simple one-dimensional Bjorken flow \cite{Bjorken:1982qr}. Eqs~{\eqref{eq:Wex}}-{\eqref{eq:Wbeta}} are the main results of this work, generalizing Maxwell's equations on top of a Bjorken flow. Later, we use the mode decomposition of Green's function to solve these nonlinear coupled wave equations, with  Eq.\eqref{Eq:Greensol} being the solution. The present work provides exact results for the electromagnetic fields for expanding fluid without relying on any time-dependent asymptotic expansion, which is a common approach in literature. However, we make two key assumptions: (i) neglecting the influence of the electromagnetic fields on the fluid flow, and (ii) setting all dissipative quantities, such as conductivity and diffusion, to zero.
	
	The paper is arranged in the following manner: we start with a recap to the basic equations Sec.~\eqref{sec:basiceqn} followed by Sec.~\eqref{sec:BackMdel}  which decribes the background model and  the underlying assumptions, solution to the Green's function is described in  Sec.~\eqref{sec:solwaveeqn}. Finally we discuss our results in the Sec.~\eqref{sec:results} followed by conclusion and outlook in the Sec.~\eqref{sec:conclusion}.
	\section{Basic equations} {\label{sec:basiceqn}}
	We will now recap the fundamental equations that we will need later; their derivation can be found in \cite{1982MNRAS,Tsagas:2004kv}. The energy-momentum tensor of the fluid in the Landau frame reads
	\begin{equation}
		T_f^{\mu\nu}\equiv\varepsilon u^\mu u^\nu +p\Delta^{\mu\nu},
	\end{equation}
	where $\Delta^{\mu\nu}$ is the spatial projection operator defined as
	\begin{equation}
		\Delta^{\mu\nu}\equiv g^{\mu\nu}+u^\mu u^\nu\;,
	\end{equation}
	and $\varepsilon$ is the energy density, $p$ the isotropic pressure and $u^\mu$ is the fluid four-velocity respectively. As usual, the fluid four-velocity is normalized
	so that $u^\mu u_\mu=-1$.
	
	The four-velocity $u_\mu$ and projection operator $\Delta_{\mu\nu}$ can be used to decompose the covariant derivative of $u_\mu$ into irreducible basic kinematic quantitites
	\begin{equation}
		u_{\mu;\nu}=\sigma_{\mu\nu}+\omega_{\mu\nu}+\frac{\theta}{3}\Delta_{\mu\nu}-\dot{u}_\mu u_\nu\;,
	\end{equation}
	where the shear tensor $\sigma_{\mu\nu}$, the vorticity tensor $\omega_{\mu\nu}$, the expansion scalar $\theta$ and the four-accleration $\dot{u}_\mu$ are defined as
	\begin{align}
		\sigma_{\mu\nu}&\equiv\frac{\tensor{\Delta}{_\mu^\alpha}\tensor{\Delta}{_\nu^\beta}}{2}\left(u_{\alpha;\beta}+u_{\beta;\alpha}\right)-\frac{\theta}{3}\Delta_{\mu\nu}\;,\\
		\omega_{\mu\nu}&\equiv\frac{\tensor{\Delta}{_\mu^\alpha}\tensor{\Delta}{_\nu^\beta}}{2}\left(u_{\alpha;\beta}-u_{\beta;\alpha}\right)\;,\\
		\theta&\equiv \tensor{u}{^\mu_{;\mu}}\;,\\
		\dot{u}_\mu&\equiv \tensor{u}{_{\mu_;\nu}}u^\nu\;.
	\end{align}
	where $\sigma_{\mu\nu}u^\mu=\omega_{\mu\nu}u^\mu=\dot{u}^\mu u_\mu=0$ by definition.
	
	The first set of Maxwell’s equations are given as
	\begin{equation}\label{Eq:Maxwell1}
		\tensor{F}{^{\mu\nu}_{;\mu}}= -J^\nu\;,
	\end{equation}
	where $J^\nu=J^\mu_f+J^\nu_{\mathrm{ext}}$ is the total charge four-current, i.e it contains both the charged current generted in the fluid $J_f$ and also that is generated from external source $J^\nu_{\mathrm{ext}}$ e.g., from spectators. The fluid charge current obeys the conservation law
	\begin{equation}
		\tensor*{J}{_f^\mu_{;\mu}}=0.
	\end{equation}
	The second set of equations is a direct consequence of the existence of a four-potential and is given by the following relation
	\begin{equation}\label{Eq:Maxwell2}
		F_{\alpha\beta;\gamma}+ F_{\beta\gamma;\alpha}+ F_{\gamma\alpha;\beta}=0\;.
	\end{equation}
	
	As seen by an observer moving with four-velocity $u_\mu$, the electromagnetic field tensor can be decomposed into an 'electric' ($E^\mu$) and 'magnetic' ($B^\mu$) part defined by
	\begin{equation}\label{Eq:ComovingE}
		E^\mu\equiv g^{\mu\alpha} F_{\alpha\nu}u^\nu\; ,
	\end{equation}
	and 
	\begin{equation}\label{Eq:ComovingB}
		B^\mu\equiv \frac{1}{2} \epsilon^{\mu\nu\alpha\beta} F_{\alpha\beta} u_{\nu} \;,
	\end{equation}
	where $\epsilon_{\mu\nu\alpha\beta}$ is the totally antisymmetric tensor with $\epsilon_{0123}=\sqrt{-g}$.
	Using the definitions Eqs.(\ref{Eq:ComovingE}) and (\ref{Eq:ComovingB}), it can be immediately deduced that
	\begin{align}
		E_\mu u^\mu&=0\; ,\\
		B_\mu u^\mu&=0\; .
	\end{align}
	Using the above definitions, the electromagnetic tensor $F_{\mu\nu}$ can be projected onto the observer's instataneous rest frame in the following way
	\begin{equation}\label{eq:emtensor}
		F_{\mu\nu}\equiv u_\mu E_\nu-E_\mu u_\nu +\epsilon_{\mu\nu\alpha\beta}u^\alpha B^\beta .
	\end{equation}
	Similarly, the charge current can be decomposed into local charge density $\rho\equiv\rho_f=-J^\mu u_\mu$ and charge diffusion current $V^\mu\equiv V_f^\mu=J^\nu\tensor{\Delta}{_\nu^\mu}$,
	\begin{equation}
		J^\mu\equiv J^\mu_f=\rho_f u^\mu + V_f^\mu\;.
	\end{equation}
	Note that as mentioned in the introduction we will not consider any external charges and currents here, all of them are of fluid origin, and for later convenience we will drop the subscript '$f$'. 
	
	Maxwell's equations Eq.(\ref{Eq:Maxwell1}) and Eq.(\ref{Eq:Maxwell2}) can be decomposed into temporal and spatial parts using $u_\mu$ and the projector $\Delta_{\mu\nu}$ yielding the following set of equations
	\begin{align} \label{eq:echargedensity}
		\tensor{E}{^\mu_{;\mu}} -B^{\mu\nu}\omega_{\mu\nu}- E^\nu \dot{u}_\nu&=\rho\;,\\
		\tensor{B}{^\mu_{\nu;\mu}}-u_\nu B_{\alpha\beta}\omega^{\alpha\beta} - E^\mu\left(\omega_{\mu\nu}+\sigma_{\mu\nu}-\frac{2\theta}{3}\Delta_{\mu\nu} \right) + \tensor{\Delta}{_\nu^\alpha}\dot{E}_\alpha&=-V_{\nu}\;,\\
		\tensor{B}{^\mu_{;\mu}} -E^{\mu\nu}\omega_{\mu\nu}- B^\nu \dot{u}_\nu&=0\;,\\ \label{eq:bchargedensity}
		\tensor{E}{^\mu_{\nu;\mu}}-u_\nu E_{\alpha\beta}\omega^{\alpha\beta} - B^\mu\left(\omega_{\mu\nu}+\sigma_{\mu\nu}-\frac{2\theta}{3}\Delta_{\mu\nu} \right) + \tensor{\Delta}{_\nu^\alpha}\dot{B}_\alpha&=0\;,
	\end{align}
	where we have defined new antisymmetric tensors $B_{\mu\nu}\equiv \epsilon_{\mu\nu\alpha\beta}u^\alpha B^\beta$, $E_{\mu\nu}\equiv \epsilon_{\mu\nu\alpha\beta}u^\alpha E^\beta$.
	
	The electromagnetic energy-momentum tensor is given as
	\begin{equation}
		T^{\mu\nu}_{em}\equiv F^{\lambda\mu}\tensor{F}{_\lambda^\nu}-\frac{1}{4}g^{\mu\nu}F_{\alpha\beta}F^{\alpha\beta}\; ,
	\end{equation}
	and the total energy-momentum tensor of the system is given by
	\begin{equation}
		T^{\mu\nu}\equiv T_f^{\mu\nu}+T^{\mu\nu}_{em}\; .
	\end{equation}
	The divergence of $T^{\mu\nu}$ is given as
	\begin{equation}
		\tensor{T}{^{\mu\nu}_{;\mu}}=0\;.
	\end{equation}
		These equations also imply that
		\begin{equation}\label{eq:partialTmunuf}
			T^{\mu\nu}_{f;\mu}=F^{\nu\lambda}J_{f,_{\lambda}}.
		\end{equation}
		Projecting Eq.\eqref{eq:partialTmunuf} in the direction of $u_\nu$ and $\tensor{\Delta}{^\alpha_\nu}$  gives the following equations of motion
		\begin{align}
			\dot{\varepsilon} &=-((\varepsilon+p)\theta + E^{\lambda}V_{\lambda})\label{eq:redenrdnsty}\\
			\dot{u}^\alpha&=\frac{1}{(\varepsilon+p)}\left(\nabla^\alpha p+E^\alpha \rho-B^{\alpha\lambda}V_{{\lambda}}\right)\label{eq:redmdnsty}
		\end{align}	
		The second and third terms of Eq.\eqref{eq:redmdnsty} are the Lorentz forces which do work on the fluid. Here, we have neglected dissipative forces arising from shear and bulk viscosity etc. Eqs.\eqref{eq:echargedensity}-\eqref{eq:bchargedensity} and Eqs.\eqref{eq:redenrdnsty}-\eqref{eq:redmdnsty}, together with an equation of state for the fluid  completely describe the system under consideration provided consistent initial and boundary data are given. In the next section, we will simplify these equations by ignoring the back-reaction of electromagnetic fields on the fluid  described in Eqs.\eqref{eq:redenrdnsty}-\eqref{eq:redmdnsty}. 	
	
	\section{Background model}\label{sec:BackMdel}
	The previous analysis is quite general given that the background evolution is specified, Eqs.{\eqref{eq:echargedensity}}-{\eqref{eq:bchargedensity}} apply to a range of physical situations. For example, in the absence of matter sources one can always set the
	observer’s acceleration to zero. Similarly, if the fluid evolution is stationary and non-rotating we can set expansion scalar and vorticity tensor to zero, etc. In the following we will assume that the background fluid is undergoing a longitudinal boost-invariant Bjorken expansion  \cite{Bjorken:1982qr}. As well known the invariance under boosts is easily manifest by using Milne coordinates, than the Minkowski coordinates, we will be using the former. 
	
	The line element in Milne coordinates $(\tau,x,y,\eta)$ is given as:
	\begin{equation}\label{Eq:LineElement}
		ds^2 =-d\tau^2 + dx^2 + dy^2 + \tau^2 d\eta^2
	\end{equation}
	Eq.\eqref{Eq:LineElement} is manifestly invariant under the following combined symmetry $SO(1, 1)\otimes ISO(2)\otimes Z_2$, namely boost-invariance along the beam direction $\eta$, rotational and translational invariance in the transverse $(x, y)$ coordinates and reflection under $\eta \rightarrow -\eta$. The flow consistent with the combined symmetry property is $u^{\mu} = \left( 1,0,0,0\right)$. Similarly, the only nonvanishing Christoffel symbols are: $\Gamma^{\tau}_{\eta\eta}\equiv\tau$, $\Gamma^{\eta}_{\tau\eta}=\Gamma^{\eta}_{\eta\tau}\equiv 1/\tau$. We also observe that Bjorken symmetry implies $\omega_{\mu\nu}=\dot{u}^\mu\equiv 0$, $\theta\equiv1/\tau$ and
	$\sigma_{\mu\nu}\equiv$ diag$\left( 0, -1/(3\tau), -1/(3\tau) , 2\tau/3 \right)$.

		However, it should be noted that the generic anisotropy of the electromagnetic energy-momentum tensor causes the Maxwell's fields to be incompatible with the high symmetry of the Milne metric. When considering ideal Bjorken flow- that is, in the absence of dissipation and effects of the electromagnetic field- Eq.\eqref{eq:redmdnsty} becomes trivially zero. This is attributed to the expansion being boost invariant, thereby resulting in zero acceleration. In a similar vein, Eq.\eqref{eq:redenrdnsty} yields the well-understood Bjorken scaling $\varepsilon\sim \tau^{-4/3}$ assuming squared speed of sound $c_s^2=1/3$. If there exists a back-reaction from the electromagnetic field impinging on the fluid, this will induce net acceleration. Such acceleration will subsequently influence the electromagnetic fields as per  Eqs.\eqref{eq:echargedensity}-\eqref{eq:bchargedensity}. Moreover, the influence of the electric field will lead to modifications in Eq.\eqref{eq:redenrdnsty}, altering the conventional Bjorken scaling $\varepsilon\sim \tau^{-4/3}$. For instance, in \cite{Dash:2022xkz}, the authors utilized a 1+1-dimensional transversely homogeneous resistive MHD calculation without ignoring any back-reaction to demonstrate that boost invariance is broken due to the ensuing self-consistent dynamics of matter and electromagnetic fields.

		Central to our discussion is an assumption. Given that the right side of Eq.\eqref{eq:redmdnsty}, containing the Lorentz forces, is multiplied by the factor $1/(\varepsilon+P)$, we can introduce the inverse plasma $\beta$-parameter: $\beta^{-1} \equiv B_0^2/(2p_0)$ responsible for determining the relative strength. If the inverse plasma $\beta$-parameter $\beta^{-1}\ll 1$, we can ignore the back-reaction of the electromagnetic field on the fluid. Since it is known that the strength of the electromagnetic field (that produced by spectators in mid-central collision) decreases faster with an increase of collision energy, while the energy density of the fluid is comparably higher with an increase in collision energy, it is expected that the $\beta^{-1}$ is small at large collision energy. Nevertheless, it could be possible that certain regions of the fireball, for example, the periphery of the fireball can have large $\beta^{-1}$ even at moderate energies.  In the remainder of this section, we will work in this regime and neglect any back-reaction of the electromagnetic fields on the background fluid.

	Lastly, we assume that the fluid is an ideal insulator with vanishing conductivity and, hence, zero diffusion current $V^\mu_f$ according to Ohm's law. This assumption is also not too bold since the conductivity of the plasma obtained from lattice simulations is small, with $\sigma/T= 8 \pi \alpha_{\mathrm{EM}}/3 \simeq 0.06$ \cite{Aarts:2020dda}, where $\sigma$ is the conductivity, $T$ is the temperature, and $\alpha_{\mathrm{EM}}$ is the fine structure constant. However, for the sake of brevity, we keep this term in the following derivation but drop it later when we discuss specific cases (see Sec.~\eqref{sec:results}).

	With the above assumption, Eqs.{\eqref{eq:echargedensity}}-{\eqref{eq:bchargedensity}} simplify to: 
	
	\begin{eqnarray} \label{eq:gauss}
		\partial_x{E_x}+\partial_y{E_y}+\tau^{-2}\partial_\eta{E_\eta}&=&\rho\;,\\
		\partial_x{B_x}+\partial_y{B_y}+\tau^{-2}\partial_\eta{B_\eta}&=&0\;,
	\end{eqnarray}
	which  are usual Gauss law for electric and magnetic fields. Similarly the  equations for Faraday’s law can be given as:
	\begin{eqnarray}
		\partial_\tau\left(\tau B_x\right)&=&-(\partial_y E_\eta-\partial_\eta E_y)\;,\\
		\partial_\tau\left(\tau B_y\right)&=&(\partial_x E_\eta-\partial_\eta E_x)\;,\\
		\partial_\tau\left(\tau^{-1} B_\eta\right)&=&-(\partial_x E_y-\partial_y E_x)\;,
	\end{eqnarray}
	and equations for Ampère’s law are given as :
	\begin{eqnarray}
		\partial_\tau\left(\tau E_x\right)&=&(\partial_y B_\eta-\partial_\eta B_y)-\tau V^x\;,\\
		\partial_\tau\left(\tau E_y\right)&=&-(\partial_x B_\eta-\partial_\eta B_x)-\tau V^y\;,\\ \label{eq:ampere}
		\partial_\tau\left(\tau^{-1} E_\eta\right)&=&(\partial_x B_y-\partial_y B_x)-\tau V^\eta\;,\label{eq:Ampere}
	\end{eqnarray}
	As mentioned in the introduction, we are interested in finding the wave equation of electromagnetic fields in this expanding background. To realize the former, we first redefine the electromagnetic fields along with the charge and currents in the following way: $\tilde{E}_{(x,y)}=\tau E_{(x,y)},~\tilde{B}_{(x,y)}=\tau B_{(x,y)}~\tilde{E}_\eta= \tau^{-1} E_\eta~,\tilde{B_\eta}= \tau^{-1} B_\eta,$ and $\tilde{\rho}= \tau \rho,~\tilde{V}^i= \tau^2 V^i$. 
	
	With the following redefinations, the Maxwell's equation Eq.\eqref{eq:gauss}-\eqref{eq:Ampere} simplify to :
	\begin{eqnarray} \label{eq:1}
		\partial_i\tilde{E_i}&=&\tilde{\rho}\;,\\ {\label{eq:bi}}
		\partial_i\tilde{B_i}&=&0\;, \\ {\label{eq:bx}}
		\partial_\tau\tilde{B}_x&=&\tau^{-1}\partial_\eta \tilde{E_y}-\tau\partial_y \tilde{E_\eta}\;,\\  {\label{eq:by}}
		\partial_\tau\tilde{B}_y&=&\tau\partial_x \tilde{E_\eta}-\tau^{-1}\partial_\eta \tilde{E_x}\;,\\  {\label{eq:beta}}
		\partial_\tau\tilde{B}_\eta&=&\tau^{-1}\partial_y \tilde{E_x}-\tau^{-1}\partial_x \tilde{E_y}\;, \\  {\label{eq:ex}}
		\partial_\tau\tilde{E}_x&=&\tau\partial_y \tilde{B_\eta}-\tau^{-1}\partial_\eta \tilde{B_y}- \tau^{-1}\tilde{V}^x\;,\\  {\label{eq:ey}}
		\partial_\tau\tilde{E}_y&=&\tau^{-1}\partial_\eta \tilde{B_x}-\tau\partial_x \tilde{B_\eta}- \tau^{-1}\tilde{V}^y\;,\\ {\label{eq:8}}
		\partial_\tau\tilde{E}_\eta&=&\tau^{-1}\partial_x \tilde{B_y}-\tau^{-1}\partial_y \tilde{B_x}- \tau^{-1}\tilde{V}^\eta \;,
	\end{eqnarray}
	Now, to get the wave equation we employ the following procedure, e.g, to get the wave equation for $\tilde{B}_x$, we take the $\partial_{\eta}$(Eq.\eqref{eq:ey}) and $\partial_{y}$(Eq.\eqref{eq:8}) and substitute the former in $\partial_{\tau}$(Eq.\eqref{eq:bx}) while making use of the constraints Eq.\eqref{eq:1} and Eq.\eqref{eq:bi}. Similar procedure is carried for other components of electromagnetic fields and we have the following
	\begin{align} 
		\Box\tilde{E}_x&=2\partial_y{\tilde{B}_\eta}-(\partial_x\tilde{\rho}+ \tau^{-1}\partial_\tau\tilde{V}^x)\;,{\label{eq:Wex}}\\
		\Box\tilde{E}_y&=-2\partial_x{\tilde{B}_\eta}-(\partial_y\tilde{\rho}+ \tau^{-1}\partial_\tau\tilde{V}^y)\;,{\label{eq:Wey}}\\
		\Box\tilde{E}_\eta&=-\tau^{-1}\partial_\tau \tilde{V}^{\eta}-\tau^{-2}\partial_\eta \tilde{\rho}\;,{\label{eq:Weeta}}\\
		\Box\tilde{B}_x&=-2\partial_y{\tilde{E}_\eta}+\partial_y \tilde{V}^\eta-\tau^{-2}\partial_\eta\tilde{V}^y\;,{\label{eq:WBx}}\\
		\Box\tilde{B}_y&=2\partial_x{\tilde{E}_\eta}-\partial_x \tilde{V}^\eta+\tau^{-2}\partial_\eta\tilde{V}^x\;,{\label{eq:WBy}}\\ 
		\Box\tilde{B_\eta}&=\tau^{-2}\partial_x \tilde{V}^{y}-\tau^{-2}\partial_y \tilde{V}^{x}\;{\label{eq:Wbeta}}, 
	\end{align}
	where $\Box$ is the d'Alembert operator in the Milne coordinates, defined as :
	\begin{equation}\label{eq:dalembert}
		\Box\equiv  \partial^2_{\tau}+\tau^{-1}\partial_{\tau}-\tau^{-2}\partial_\eta^2 -\partial_x^2-\partial_y^2.
	\end{equation}
	Eqs.\eqref{eq:Wex}-\eqref{eq:Wbeta} are the main results of this work. These equations, without the external charges and currents, can be compared to the standard source-free wave equation in Minkowski coordinates~\cite{Gursoy:2018yai}, which does not have the additional couplings between the field components appearing in the right-hand side as seen in the former coordinate system. One interesting consequence of these coupling terms is that one can produce a magnetic field for a stationary charge in an expanding medium without even having any charge current $\tilde{V}^i$. The resulting magnetic fields are dictated by the gradients of electric fields, which act as sources\footnote{The solutions to these equations in Minkowski coordinates are sometimes also called Jefimenko's equations or Jefimenko-Feynman formula \cite{Griffiths:1492149}.}. This will be discussed briefly in Sec.~\eqref{sec:results}. The origin of these terms can be traced back to the non-vanishing expansion scalar $\theta$ and shear stress tensor $\sigma^{\mu\nu}$ in Eqs.\eqref{eq:echargedensity}-\eqref{eq:bchargedensity}. We must also stress that the electromagnetic fields obtained from the solutions of these Eqs.\eqref{eq:Wex}-\eqref{eq:Wbeta} are not the coordinate-transformed solution of electromagnetic fields in Minkowski coordinates. Unless one solves Eqs.\eqref{eq:Wex}-\eqref{eq:Wbeta} with longitudinal fluid velocity $v^z\equiv u^z/u^t=\tanh\eta$ in the latter coordinate, the solutions will differ (here $u^z$ and $u^t$ are the components of the four-velocity in Minkowski coordinates). Since the velocity in Minkowski coordinates is coordinate-dependent, the fields cannot be obtained simply by boosting from the rest frame to this frame. In the following section, we shall solve Eqs.\eqref{eq:Wex}-\eqref{eq:Wbeta} based on mode decomposition of Green's equation.

	\section{Solution of the wave equations}\label{sec:solwaveeqn}
	To solve the non-linear coupled wave equation system Eqs.\eqref{eq:Wex}-\eqref{eq:Wbeta}, we first notice that the longitudinal components of the electromagnetic fields solely depend on external sources while the transverse components are dependent on the gradient of the former. Therefore, we can solve the system of equations iteratively, by first solving for the longitudinal components and then using this solution to find the transverse components. The equation for the longitudinal components is a well-known problem in the literature \cite{nariai1976propagators,Padmanabhan:1990fm,Rindori:2021quq}, and is encountered in solving the Klein-Gordon equation in Milne coordinates. We mention that often either the WKB procedure \cite{Sagnotti:1981wq,SVDhurandhar_1984} or the mode decomposition of Green's function \cite{Burko:2002ge,poisson2011motion} is used to solve these equations, but we adopt the latter approach which leads to an exact solution. For completeness, we outline our calculations in the paper.

	Using the 2-point Green function $G(x_\mu;x'_\mu)$ between $x_\mu\equiv (\tau,\mathbf{x}_\bot,\eta)$ and $x'_\mu\equiv (\tau',\mathbf{x}^\prime_\bot,\eta ')$ we make the following definitions
	\begin{align}
		G_r(x_\mu;x'_\mu)&=-\Theta(\tau-\tau')G(x_\mu;x'_\mu)\:,\\
		G_a(x_\mu;x'_\mu)&=\Theta(\tau'-\tau)G(x_\mu;x'_\mu)\:,\\ 
		\bar{G}(x_\mu;x'_\mu)&=\frac{1}{2}\left[G_r(x_\mu;x'_\mu)+G_a(x_\mu;x'_\mu)\right]\:,\label{eq:Symmpropdefn}
	\end{align}
	where $\Theta(\tau-\tau^{'})$ is the Heaviside step function with $G_r(x_\mu;x'_\mu)$, $G_a(x_\mu;x'_\mu)$ and $\bar{G}(x_\mu;x'_\mu)$ being the retarded, advanced and symmetric propagators respectively. 
	
	Now we decompose the Green function into modes via a Fourier expansion 
	\begin{equation}
		G(x_\mu;x'_\mu)=\frac{1}{(2\pi)^3\sqrt{\tau\tau'}}\int d^3k e^{i[\mathbf{k}_\bot\cdot (\mathbf{x}_\bot-\mathbf{x}^\prime_\bot)+k_{\eta} (\eta-\eta^{\prime})]}\left[a_{k_{\eta}}(\tau)b_{k_{\eta}}(\tau')-a_{k_{\eta}}(\tau')b_{k_{\eta}}(\tau)\right]\:,\label{Eq:GreenDecom}
	\end{equation}
	where $p_\mu\equiv(\mathbf{k}_\bot,k_\eta)$. If we denote the particular solutions of the above equations by $a_{k_{\eta}}(\tau)$ and $b_{k_{\eta}}(\tau)$ in such a way that they satisfy the following relations
	\begin{align}
		a_{k_{\eta}}(\tau)&=1\:,& \partial_\tau a_{k_{\eta}} (\tau)&=0\:,\\
		b_{k_{\eta}}(\tau)&=0\:,& \partial_\tau b_{k_{\eta}}(\tau)&=1\:,
	\end{align}
	at a given instant $\tau=1$, then it can be verified that following relation is valid for  any $\tau$, 
	\begin{equation}
		a_{k_{\eta}}(\tau)\partial_\tau b_{k_{\eta}} (\tau)-\partial_\tau a_{k_{\eta}} (\tau) b_{k_{\eta}}(\tau)=1.
	\end{equation}
	given the Green's function $G(x_\mu;x'_\mu)$ satisfies the homogeneous wave equation
	\begin{equation}
		\Box G(x_\mu;x'_\mu)=0
	\end{equation}
	together with boundary condition:
	\begin{align}
		G(x_\mu;x'_\mu)&=0,&
		\partial_\tau G(x_\mu;x'_\mu)=-\frac{1}{\tau}\delta(x_i-x'_i)\;,
	\end{align}
	at $\tau=\tau^\prime$.
	
	We can also show that the symmetric propagator $\bar{G}(x_\mu;x'_\mu)$ satisfies the following inhomgeneous wave equation,
	\begin{equation}
		\Box\bar{G}(x_\mu;x'_\mu)=\frac{1}{\sqrt{\tau\tau '}}\delta^4(x_\mu-x'_\mu)
	\end{equation}
	where $\delta^4(x_\mu-x'_\mu)=\delta(\tau-\tau ')\delta^3(x_i-x'_i)$. Now using the definition of d'Alembert operator from Eq.\eqref{eq:dalembert}, it can be verified that the particular solution satisfy following relations, 
	\begin{equation}
		a_{k_{\eta}}(\tau)b_{k_{\eta}}(\tau ')- a_{k_{\eta}}(\tau ') b_{k_{\eta}}(\tau)= \frac{\pi \sqrt{\tau\tau '}}{2}\left[J_{ik_{\eta}}\left(k_\bot \tau\right)Y_{ik_{\eta}}\left(k_\bot \tau '\right)-J_{ik_{\eta}}\left(k_\bot \tau '\right)Y_{ik_{\eta}}\left(k_\bot \tau \right)\right]\;,\label{Eq:modefunc}
	\end{equation}
	where the $J_{ik_{\eta}}\left(k_\bot \tau\right)$, $Y_{ik_{\eta}}\left(k_\bot \tau\right)$ are the Bessel function of $1^{\mathrm{st}}$ and $2^{\mathrm{nd}}$ kind respectively with $k_\bot\equiv|\mathbf{k}_\bot|$. Plugging Eq.\eqref{Eq:modefunc} into Eq.\eqref{Eq:GreenDecom}, we arrive at the following equation for Green's function:
	\begin{align} \nonumber
		G(x_\mu;x'_\mu)=&\frac{1}{16\pi^2}\int_{-\infty}^{\infty} dk_{\eta} e^{ik_{\eta}(\eta-\eta')}\int_0^\infty dk_\bot k_\bot\int_0^{2\pi} d\theta e^{ik_\bot r_\bot\cos\theta}\cdot\\ \nonumber
		&\left[J_{ik_{\eta}}\left(k_\bot \tau\right)Y_{ik_{\eta}}\left(k_\bot \tau '\right)-J_{ik_{\eta}}\left(k_\bot \tau '\right)Y_{ik_{\eta}}\left(k_\bot \tau \right)\right]\nonumber\\ \nonumber
		=&\frac{1}{8\pi}\int_{-\infty}^{\infty} dk_{\eta} e^{ik_{\eta}(\eta-\eta')}\int_0^\infty dk_\bot k_\bot J_0(k_\bot r_\bot)\cdot\\ \nonumber
		&\left[J_{ik_{\eta}}\left(k_\bot \tau\right)Y_{ik_{\eta}}\left(k_\bot \tau '\right)-J_{ik_{\eta}}\left(k_\bot \tau '\right)Y_{ik_{\eta}}\left(k_\bot \tau \right)\right]\\ \nonumber
		=&-\frac{\epsilon(\tau-\tau ')\theta(\lambda^2)}{4\pi}\int_0^\infty k_\bot dk_\bot  J_0(\lambda k_\bot)J_0(r_\bot k_\bot)\\ {\label{eq:solG}}
		=&-\frac{\epsilon(\tau-\tau ')}{2\pi}\delta(s^2)
	\end{align}
	where $\lambda^2=\tau^2+\tau '^2-2\tau\tau '\cosh (\eta-\eta')$, $r^2_\bot\equiv (x-x')^2 + (y-y')^2$, $s^2=\lambda^2-r_\bot^2$ and  $\epsilon(\tau-\tau ')=\Theta(\tau-\tau ')-\Theta(\tau'-\tau )$ respectively.
	
	Substituting Eq.\eqref{eq:solG} into the definition of symmetric propagator Eq.\eqref{eq:Symmpropdefn}, we have
	\begin{equation}{\label{eq:solGbar}}
		\bar{G}(x_\mu;x'_\mu)=\frac{1}{4\pi}\delta(\tau^2+\tau '^2-2\tau\tau '\cosh (\eta-\eta')-r_\bot^2).
	\end{equation}
	Eq.\eqref{eq:solG} is very similar to its counterpart in the usual representation in Minkowski space-time but the expression for $s^2$, in Eq.\eqref{eq:solG} ($s^2=0$ has support only at the lightcone) is entirely different from that in the later.
	
	Thus, for any field $\Phi(x_{\mu})$ satisfying an equation of the form.
	\begin{equation}
		\Box \Phi (x_\mu)= S (x_\mu)
	\end{equation}
	with a generic source $S(x_\mu)$, we have the solution of the form :
	\begin{equation}\label{Eq:Greensol}
		\Phi(x_\mu)=\int \bar{G}(x_\mu; x'_\mu) S(x_\mu') \sqrt{g^\prime}d^4 x'
	\end{equation}

	\section{Results} \label{sec:results}
	In this section, we present the results for the electromagnetic fields generated by the participants. First, we consider a simple case of a stationary point charge, for which the fields corresponds to the fields from the Liénard–Wiechert potential for the expanding fluid. This is to test our formulation for a simplistic case. Next we move to a more realistic senario of charge particles being distributed in a gaussian profile in the transverse plane with a constrain on the region of participant charge distribution.

	\subsection{Field of a stationary point charge} \label{sec:pointcharge}

	Firstly, the charge density of a point particle at rest (co-moving) in an expanding fluid with four velocity $u^{\mu} = \left( 1,0,0,0\right)$, is given as
	\begin{equation}\label{Eq:pointchrg}
		\rho(\tau,\mathbf{x})=Ze\frac{\delta^3(\mathbf{x}-\mathbf{x}_0)\Theta(\tau-\tau_0)}{\tau}
	\end{equation}
	where $Ze$, $\mathbf{x}_0$, is the magnitude and position of the charge. To avoid the singularity of the Green's function at $\tau=0$, we assume that the charge appeared at a finite time in the past, $\tau=\tau_0$. This is motivated by the fact that hydrodynamics in heavy-ion collisions typically starts after a finite time, around $\sim 0.5-0.6$ fm. It should be noted that, in this theory, there is no conservation law for charge, and a charge can be spontaneously created if there is enough energy available. Additionally, since we have assumed the fluid to be a perfect insulator, the charge diffusion current $V^\mu$ is zero. Therefore, the particular solution for the $\eta$ component of the magnetic field Eq.\eqref{eq:Wbeta} can be set to zero without loss of generality, and it decouples from the other components of the electromagnetic fields. By substituting the gradient of the point charge Eq.\eqref{Eq:pointchrg} as a source in the Green's function Eq.\eqref{Eq:Greensol}, we obtain
	\begin{equation}\label{Eq:EetaInt}
		\tilde{E}_\eta(\tau ,\mathbf{x})=-Ze\int \frac{1}{{(\tau^\prime)}^2}\bar{G}(\tau ,\mathbf{x};\tau^\prime ,\mathbf{x}^\prime)\partial_\eta \tilde{\rho}(\tau^\prime ,\mathbf{x}^\prime)\sqrt{g(\tau^\prime)}d^3\mathbf{x}^\prime d\tau^\prime
	\end{equation}
	as a solution of Eq.\eqref{eq:Weeta}. Integration of Eq.\eqref{Eq:EetaInt} with the symmetric Green's function of Eq.\eqref{eq:solGbar} is elementary, and we obtain
	\begin{equation}\label{Eq:Eetasol}
		\tilde{E}_\eta(\tau ,\mathbf{x})= 
		\begin{cases}
			Ze \frac{\tau\sinh(\eta-\eta_0)}{4\pi{[(r_\bot-r_{\bot 0})^2 +\tau^2\sinh^2(\eta-\eta_0)]}^{3/2}},& \text{if } \tau_0<\tau_f(\mathbf{x};\mathbf{x}_0)<\tau\\
			0,              & \text{otherwise}
		\end{cases}
	\end{equation}
	where, $\tau_f(\mathbf{x};\mathbf{x}_0):=\tau\cosh(\eta-\eta_0)-\sqrt{(r_\bot-r_{\bot 0})^2 +\tau^2\sinh^2(\eta-\eta_0)}$ and the inequality satisfies the causality constraints.
	
	The transverse components of electric fields can also be computed similarly and are given as
	\begin{equation}\label{Eq:Exsol}
		\tilde{E}_x(\tau ,\mathbf{x})=
		\begin{cases}
			Ze\frac{\tau(x-x_0)\cosh(\eta-\eta_0)}{4\pi{[(r_\bot-r_{\bot 0})^2 +\tau^2\sinh^2(\eta-\eta_0)]}^{3/2}},& \text{if } \tau_0<\tau_f(\mathbf{x};\mathbf{x}_0)<\tau\\
			0,              & \text{otherwise}
		\end{cases}
	\end{equation}
	\begin{equation}\label{Eq:Eysol}
		\tilde{E}_y(\tau ,\mathbf{x})=
		\begin{cases}
			Ze\frac{\tau(y-y_0)\cosh(\eta-\eta_0)}{4\pi{[(r_\bot-r_{\bot 0})^2 +\tau^2\sinh^2(\eta-\eta_0)]}^{3/2}},& \text{if } \tau_0<\tau_f(\mathbf{x};\mathbf{x}_0)<\tau\\
			0,              & \text{otherwise} 
		\end{cases}
	\end{equation}

	As discussed in Sec.\eqref{sec:solwaveeqn} the transverse magnetic fields can be computed by taking the gradients of longitudinal component of electric field Eq.\eqref{Eq:Eetasol}, which acts as source in Eq.\eqref{Eq:Greensol} and is given as

	\begin{equation}\label{Eq:ByInt1}
		\tilde{B}_y(\tau ,\mathbf{x})=2\int \bar{G}(\tau ,\mathbf{x};\tau^\prime ,\mathbf{x}^\prime)\partial_{x'} \tilde{E}_\eta(\tau^\prime ,\mathbf{x}^\prime)\sqrt{g(\tau^\prime)}d^3\mathbf{x}^\prime d\tau^\prime
	\end{equation}
	as a solution of Eq.\eqref{eq:WBy}. The integration over $\tau^\prime$ is elementary and we are left with
	\begin{equation}\label{Eq:ByInt}
		\tilde{B}_y(\tau ,\mathbf{x})=
		\begin{cases}
			Ze\frac{3}{8\pi^2}\int d^3 \mathbf{x}' \frac{\tau_f(\mathbf{x}';\mathbf{x})^2(x'-x_0)\sinh(\eta_0-\eta')}{{[(r'_\bot-r_{\bot 0})^2 +\tau_f(\mathbf{x}';\mathbf{x})^2\sinh^2(\eta '-\eta_0)]}^{5/2}}\frac{1}{\sqrt{(r_\bot-r_\bot ')^2 +\tau^2\sinh^2(\eta-\eta ')}},& \text{if }  (\tau_0<\tau_f(\mathbf{x}';\mathbf{x}_0)<\tau)~\land\\
			& (\tau_0<\tau_f(\mathbf{x}';\mathbf{x})<\tau)\\
			0,              & \text{otherwise} 
		\end{cases}
	\end{equation}
	where, $\tau_f(\mathbf{x}';\mathbf{x})=\tau\cosh(\eta-\eta ')-\sqrt{(r_\bot-r_\bot ')^2 +\tau^2\sinh^2(\eta-\eta ')}$. In the above expression, one of the constraints is inherited from the electric field Eq.\eqref{Eq:Eetasol} while the other is from the Green's function appearing in Eq.\eqref{Eq:ByInt1}. The $x$ component of magnetic field can also be obtained by following a similar procedure. The spatial integration in Eq.\eqref{Eq:ByInt}  can not  be reduced to an elementary form and we perform the integration numerically in the rest of this section.
	
	To gain insight into the space-time dependence of electromagnetic fields produced by the charged participants, we make a simplifying assumption that the charges are located at $\mathbf{x}_{0}=(b/2,0,0)$ and the initial time $\tau_0=0.6$~fm, where $b$ is the coordiante where the source is located taken as $b=7$~fm. We calculate the fields at points with transverse coordinates $\mathbf{x}_\bot=(0,0)$, but for variable rapidities $\eta$ and time $\tau$. The magnitude of charge $Ze$ is a free parameter, and we took $Z=79$, which is half of the total charged spectators for an Au-Au collision and is, of course, a simplification. With the above geometry, the only non-vanishing components of electromagnetic fields are $\tilde{E}_\eta,~E_x$ and $B_y$.
	
	\begin{figure}[ht] 
		\centering
		\includegraphics[width=0.45\textwidth,,height=0.25\textheight]{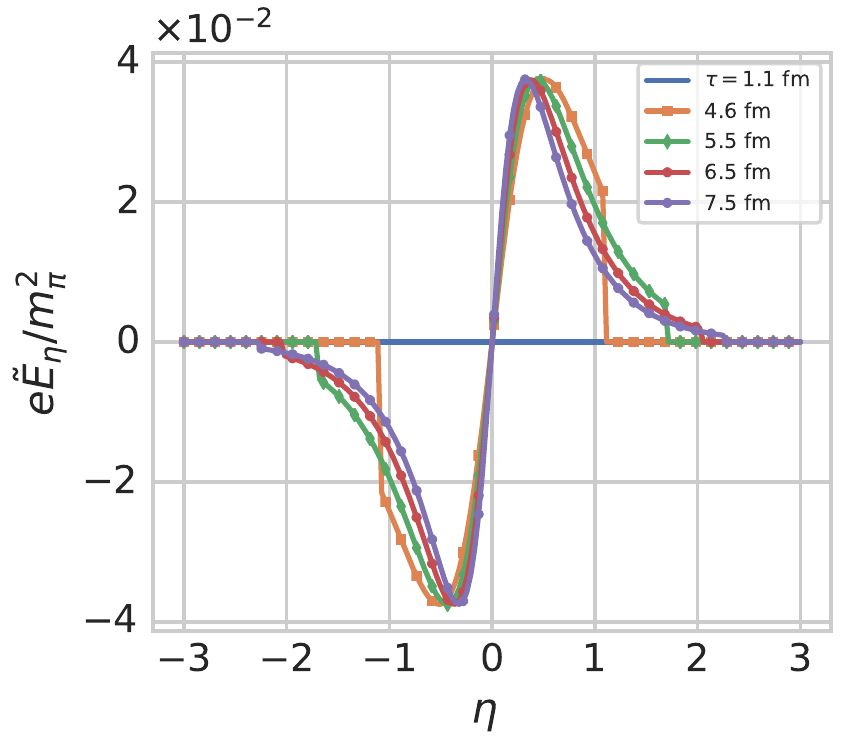} 
		\includegraphics[width=0.45\textwidth,,height=0.25\textheight]{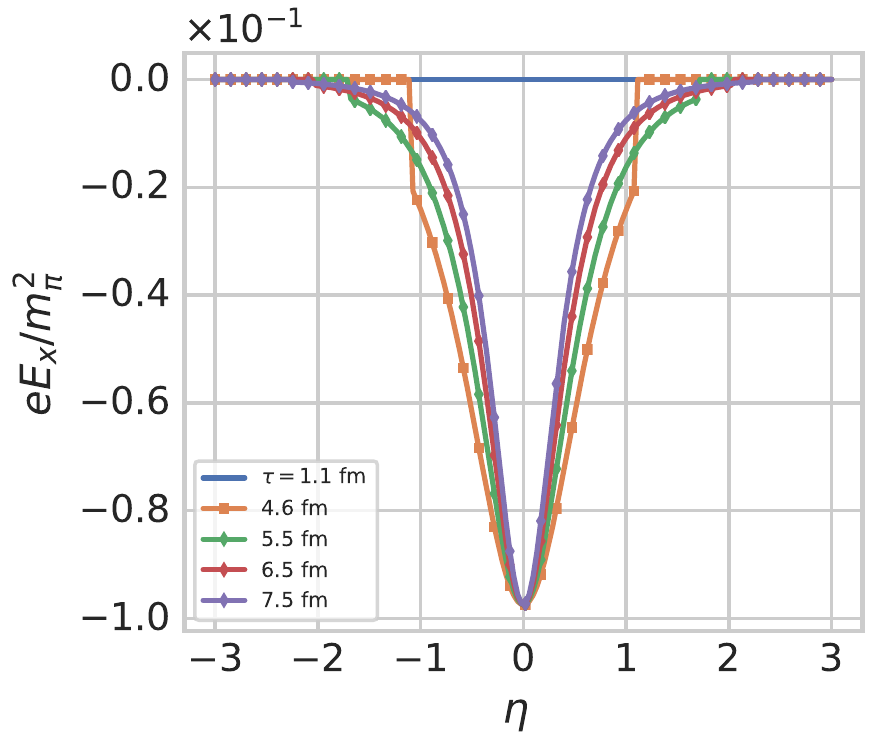} 
		\caption{Left panel: The electric-field component $e\tilde{E}_\eta$ as a function of $\eta$ for a stationary point source. Various symbols represent the values of electric field at different times. Right panel: Same as left panel but for the electric-field component $eE_x$.} \label{Electricfield_taueta}
	\end{figure}
	
	Fig.\eqref{Electricfield_taueta} shows the electric field components $e\tilde{E}_\eta$ and $eE_x$ as a function of $\eta$. The various symbols in the figure represent different time frames. As shown in the figure, the $x$-component of the electric field is even under rapidity and its magnitude is roughly ten times larger than that of the $\eta$-component, while the $\eta$-component of the electric field is odd under rapidity. At a given $\eta$, the electric fields decay as $\sim \tau^{-3}$ (see Eqs. \eqref{Eq:Eetasol} and \eqref{Eq:Exsol}). Since the fields are retarded, prior to $\tau=4.6$ fm, the electric field for both components is zero. At a later time, only the region allowed by causality experiences the electric field, which appears as a piecewise function in the figure above. This region of influence depends, of course, on the initial time $\tau_0$ and the relative distances $\mathbf{x}-\mathbf{x}_0$.     
	\begin{figure}[ht] 
		\centering
		\includegraphics[width=0.45\textwidth,,height=0.25\textheight]{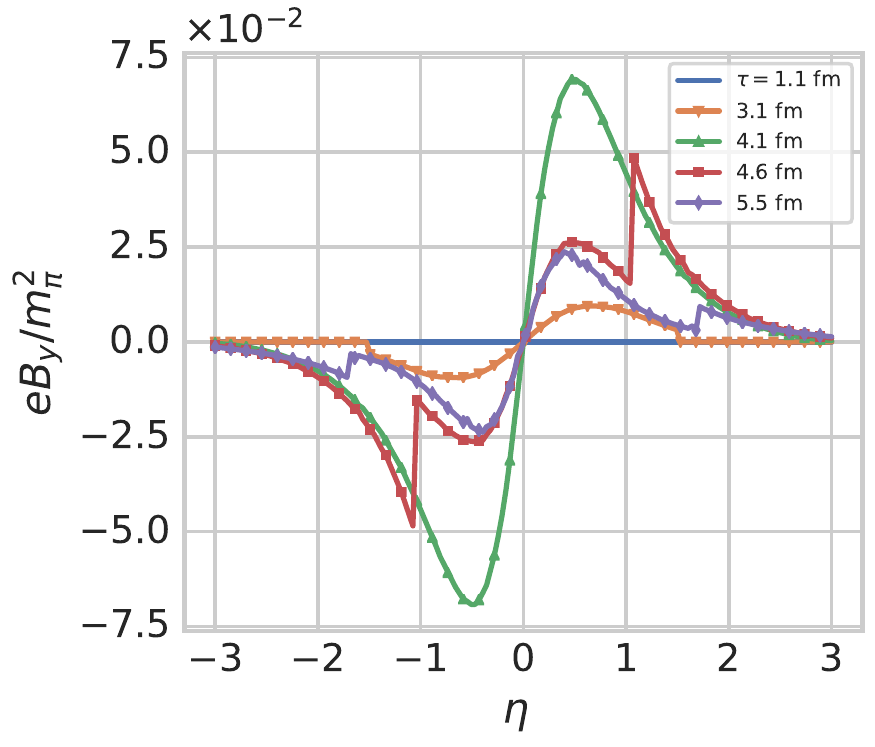} 
		\includegraphics[width=0.45\textwidth,,height=0.25\textheight]{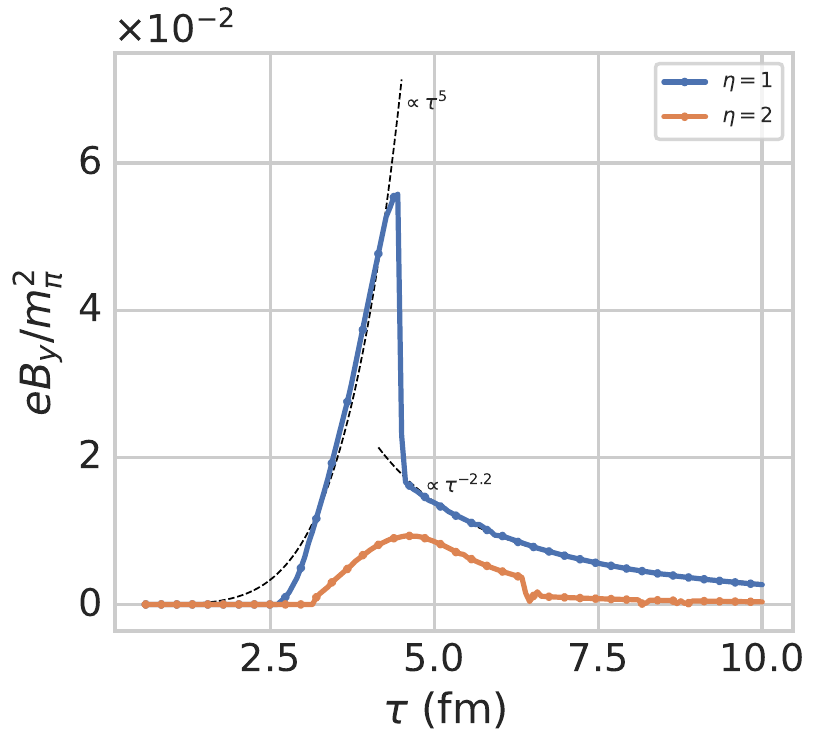} 
		\caption{Left panel: The magnetic-field component $eB_y$ as a function of $\eta$ for a stationary point source. Various symbols represent the values of magnetic field at different times. Right panel: Time evolution of $eB_y$ at rapidities $\eta=1$ and $\eta=2$ respectively. The black dotted lines are a fit to a power law function.} \label{Magneticfield_taueta}
	\end{figure}

	Fig.\eqref{Magneticfield_taueta} (left panel) shows the magnetic field component $eB_y$ as a function of $\eta$, with various symbols representing different time frames. Firstly, we note that at early times ($\tau<3.1$ fm), the magnetic field is zero, owing to the first causality constraint (see Eq. \eqref{Eq:ByInt}) inherited from the electric field. However, unlike the electric field, which has support only at $\tau_f(\mathbf{x};\mathbf{x}_0)$, the magnetic field's support $\tau_f(\mathbf{x}';\mathbf{x}_0)$ extends to a larger region of space-time. This is also readily seen in the left panel of Fig. \eqref{Magneticfield_taueta} (left panel), where the magnetic field attains non-zero values prior to the corresponding electric field, which remains zero until $\tau=4.6$ fm. Nevertheless, the magnetic field for such an early time is limited to a smaller rapidity region owing to this constraint. Next, for $3.1$ fm $<\tau<$ 4.6 fm, the magnitude of the magnetic field increases and shows continuous evolution throughout the rapidity region.
	
		The domains of influence for both the electric and magnetic fields can be discerned from their respective equations (i) electric field, $e{\tilde E}_\eta$ from Eq.~\eqref{Eq:Eetasol} given as $\tau_0<\tau_f(\mathbf{x};\mathbf{x}_0)<\tau$ and (ii) magnetic field, $e{B}_y$ from  Eq.~\eqref{Eq:ByInt} given as $\tau_0<\tau_f(\mathbf{x}';\mathbf{x}_0)<\tau~\land \tau_0<\tau_f(\mathbf{x}';\mathbf{x})<\tau$.
	\begin{figure}[ht] 
		\centering
		\includegraphics[width=0.35\textwidth,height=0.25\textheight]{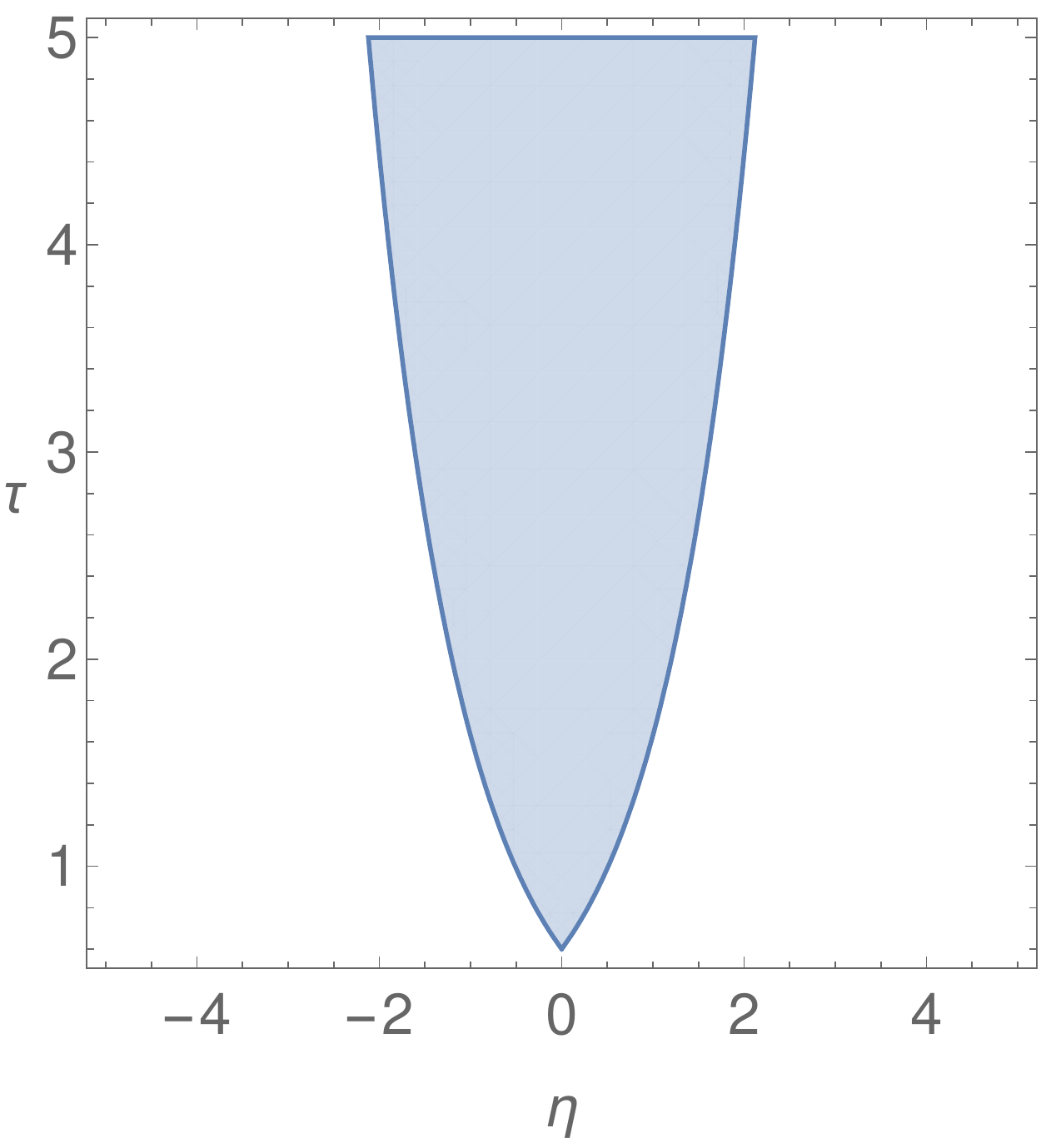} 
		\includegraphics[width=0.35\textwidth,height=0.25\textheight]{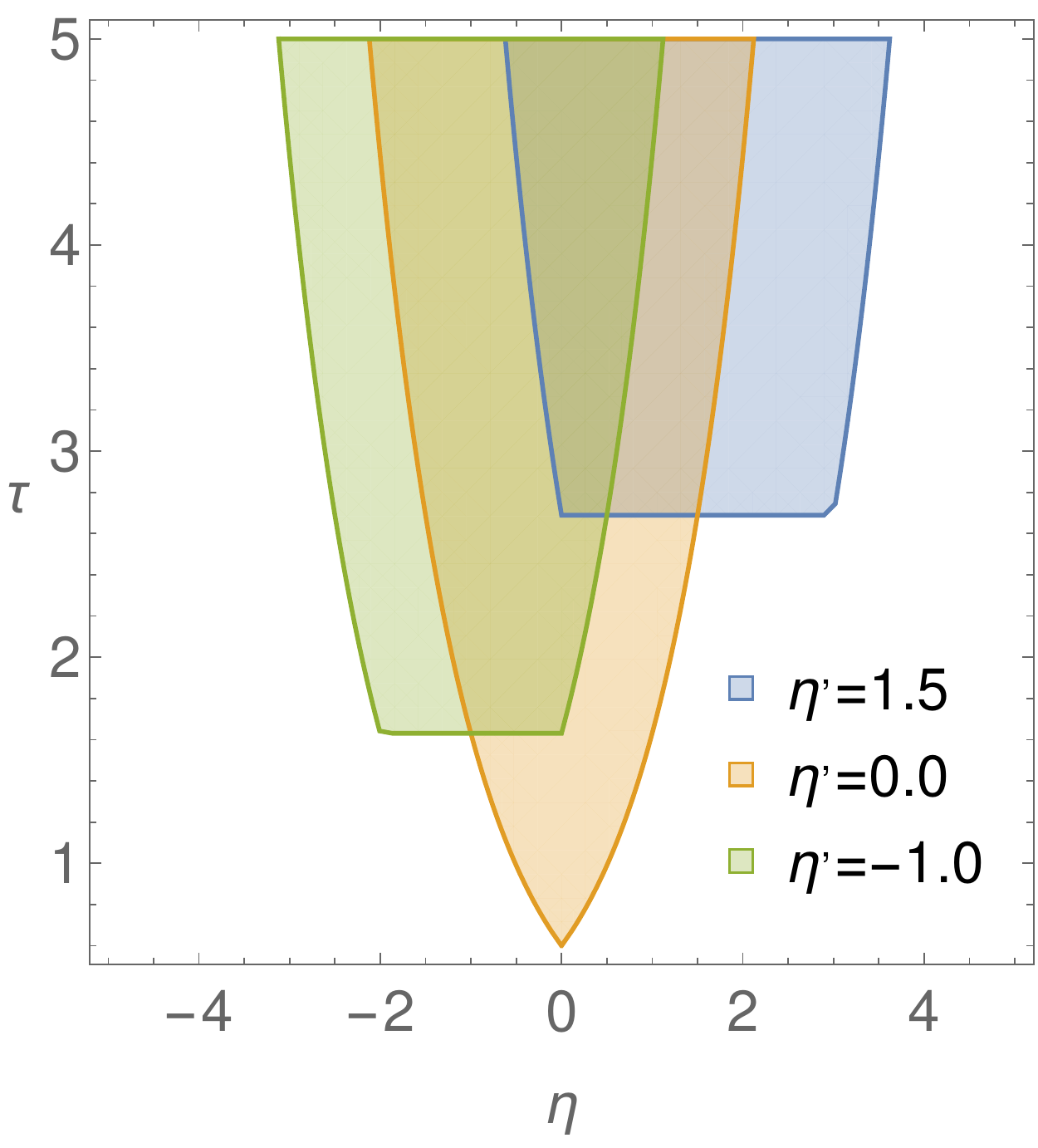} 
		\caption{Left panel: Domain of influence $\tau_0<\tau_f(\mathbf{x};\mathbf{x}_0)<\tau$ for the electric field component $e{\tilde E}_\eta$. Right panel: for the magnetic field component $eB_y$ ,$\tau_0<\tau_f(\mathbf{x}';\mathbf{x}_0)<\tau~\land \tau_0<\tau_f(\mathbf{x}';\mathbf{x})<\tau$. The transverse coordiantes and $\eta_0$ are set to zero ($r_\bot=r_{\bot 0}=\eta_0=0$) for simplicity.} \label{Domain_taueta}
	\end{figure}
	
	From Fig.\eqref{Domain_taueta} (left panel), which visualizes the domain of influence or the causal region, one can observe that for smaller values of $\tau$, the support of $\tau_f(\mathbf{x};\mathbf{x}_0)$ is restricted to narrower $\eta$ intervals for electric field. However, referring to Fig.\eqref{Domain_taueta} (right panel) for the magnetic field, the support spans larger areas based on the specific value of $\eta^\prime$. To elucidate, at $\eta^\prime=0$, the magnetic field's support mirrors that of the electric field. But when $\eta^\prime=-1.0$ or $\eta^\prime=1.5$, the depicted (shaded) region encompasses a more extensive area than the prior scenario.  As we transition to later times, both the electric and magnetic fields exhibit a progressive expansion in their supportive domains.
	
	Fig.\eqref{Magneticfield_taueta} (right panel) shows the time evolution of the magnetic field for two different values of rapidity, $\eta=1,2$ respectively. For $\eta=1$, we have also fitted the obtained numerical solution with a power law (black dotted lines). As can be seen clearly, the whole time evolution can be divided into two regions: at early times, the growth of magnetic field $\propto \tau^5$, while at late time it decays as $\propto \tau^{-2.2}$, and in the intermediate region around $\tau\sim 4.6$~fm, there is a discontinuity. For larger rapidities, e.g. $\eta=2$, the magnitude of the magnetic field is an order of magnitude smaller than for smaller rapidities, e.g. $\eta=1$. It is also instructive to compare the above results with that generated from spectators~\cite{Gursoy:2018yai}, which, at vanishing conductivity, simply decays as $\propto \tau^{-3}$, although at early times the magnitude is much larger than due to participants and depends on the colliding energy. Nevertheless, a unique feature of the magnetic field produced by the participants, unlike the spectators, is that they remain non-negligible throughout the evolution even at very late times. For example, at $\tau=10$~fm for smaller rapidity $\eta=1$, the magnitude is in the order of $\mathcal{O}(10^{-3})$ vs $\mathcal{O}(10^{-5})$ (in units of $m^2_{\pi}$) for the spectators.

	\subsection{Field of a transverse charge distribution}	
	Next, we turn to the problem of finding the fields generated by a stationary charge distribution. Without any loss of generality, we assume that the charge density of the target and projectile is distributed along the transverse plane while it is still localized in the rapidity direction. We shall make the simplifying assumption that the protons in a nucleus are uniformly distributed according to a Gaussian distribution with mean $\mathbf{x}_0$ and standard deviation $\sigma_\bot$. The charge density can then be described as follows:
	\begin{equation}\label{Eq:sourcechrg}
		\rho(\tau,\mathbf{x})=Ze\frac{f_\bot(x,x_0;y,y_0)\delta(\eta-\eta_0)}{\tau}\Theta(\tau-\tau_0)
	\end{equation}
	where $f_\bot(x,x_0;y,y_0)$ is the charge distribution in the transverse direction and which is given as:
	\begin{equation}\label{Eq:Normaldistbn}
		f_\bot(x,x_0;y,y_0)=\frac{1}{\pi\sigma_\bot^2}\left[\exp\left(-{\frac{(x-x_0)^2 +(y-y_0)^2 }{\sigma_\bot^2} }\right) + \exp\left(-{\frac{(x+x_0)^2 +(y+y_0)^2 }{\sigma_\bot^2} }\right)\right] \Theta\left(1- \frac{x^2}{r_a^2}- \frac{y^2}{r_b^2}\right)
	\end{equation}
	with $x_0=b/2$, $y_0=0$ which corresponds to the centers of the nuclei in the transverse plane and we took $\sigma_\bot =5$ fm respectively. Here $b$ is the impact parameter which is choosen as $b=7$ fm. The semi-major and semi-minor axis of the elliptical region of the participants are given by the quantities $r_a=R-x_0$ and $r_b=\sqrt{R^2-x_0^2}$, where $R$ is the radius of a nucleus and we took $R=7$ fm. The unit step function in Eq.\eqref{Eq:Normaldistbn} guarentees us that we consider only the charges in the elliptical region in the transverse plane.
	
	Following  procedure similar to point charge as discussed in Sec.~\eqref{sec:pointcharge}, we can find the different field components using the Green's function Eq.\eqref{eq:solGbar} for the equations Eqs.\eqref{eq:Wex} to~\eqref{eq:Wbeta}. The integration over $\tau$ and $\eta$ are elementary and the resulting expressions for electromagnetic field components turns out to be:
	\begin{eqnarray}\label{Eq:Eetasol1}  
		\tilde{E}_\eta(\tau ,\mathbf{x}) &=& \frac{Ze}{4\pi} \int  d^2\mathbf{x}'
		\frac{\tau\sinh(\eta-\eta_0)}{{[(x- x')^2 + (y-y')^2 +\tau^2\sinh^2(\eta-\eta_0)]}^{3/2}} f_\bot(x',x_0;y',y_0) \Theta\left(1- \frac{{x'}^2}{r_a^2}- \frac{{y'}^2}{r_b^2}\right)\\
		\tilde{B}_y(\tau ,\mathbf{x})&=& Ze\frac{3}{8\pi^2}\int  d^3\mathbf{x}' d^2\mathbf{x}'' \frac{\tau_f(\mathbf{x}';\mathbf{x})^2 (x'-x'') \sinh(\eta_0 -\eta')}{((x'-x'')^2 + (y'-y'')^2 +\tau_f(\mathbf{x}';\mathbf{x})^2 \sinh^2 (\eta_0 -\eta) )^{5/2}}\times\nonumber\\ &&\frac{f_\bot(x'',x_0;y'',y_0)}{\sqrt{(x-x')^2 + (y-y')^2 + \tau^2 \sinh^2 (\eta-\eta')}}
		\Theta\left(1- \frac{{x''}^2}{r_a^2}- \frac{{y''}^2}{r_b^2}\right)
	\end{eqnarray}
	where the integration over $\mathbf{x}'$ is again limited to the causal region satisfying the inequality $\tau_0<\tau_f(\mathbf{x}';\mathbf{x})<\tau$ along with the physical boundary of the elliptical region which is expressed via the unit-step function. Here, we have explicitly shown the expression for $\tilde{E}_\eta$ and $\tilde{B}_y$, other components of the electromagnetic fields can be calculated using a similar procedure and will not be discussed further.

	\begin{figure}[ht] 
		\centering
		\includegraphics[width=0.45\textwidth,,height=0.25\textheight]{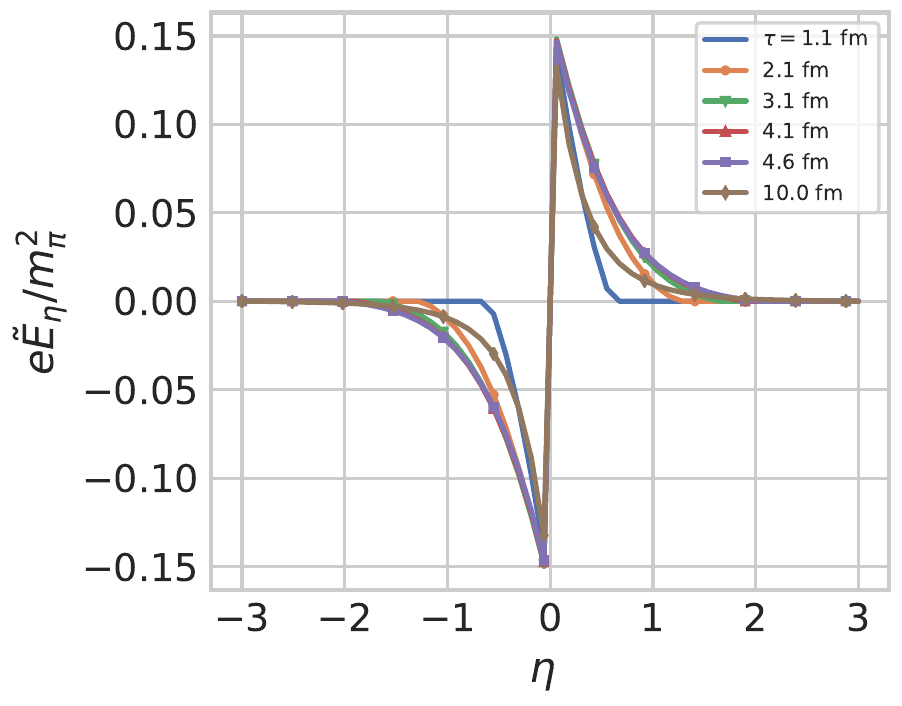} 
		\caption{The electric-field component $e\tilde{E}_\eta$ as a function of $\eta$ . Various symbols represent the values of electric field at different times.} \label{fig:Eeta1}
	\end{figure}

	\begin{figure}[ht] 
		\centering
		\includegraphics[width=0.45\textwidth,,height=0.25\textheight]{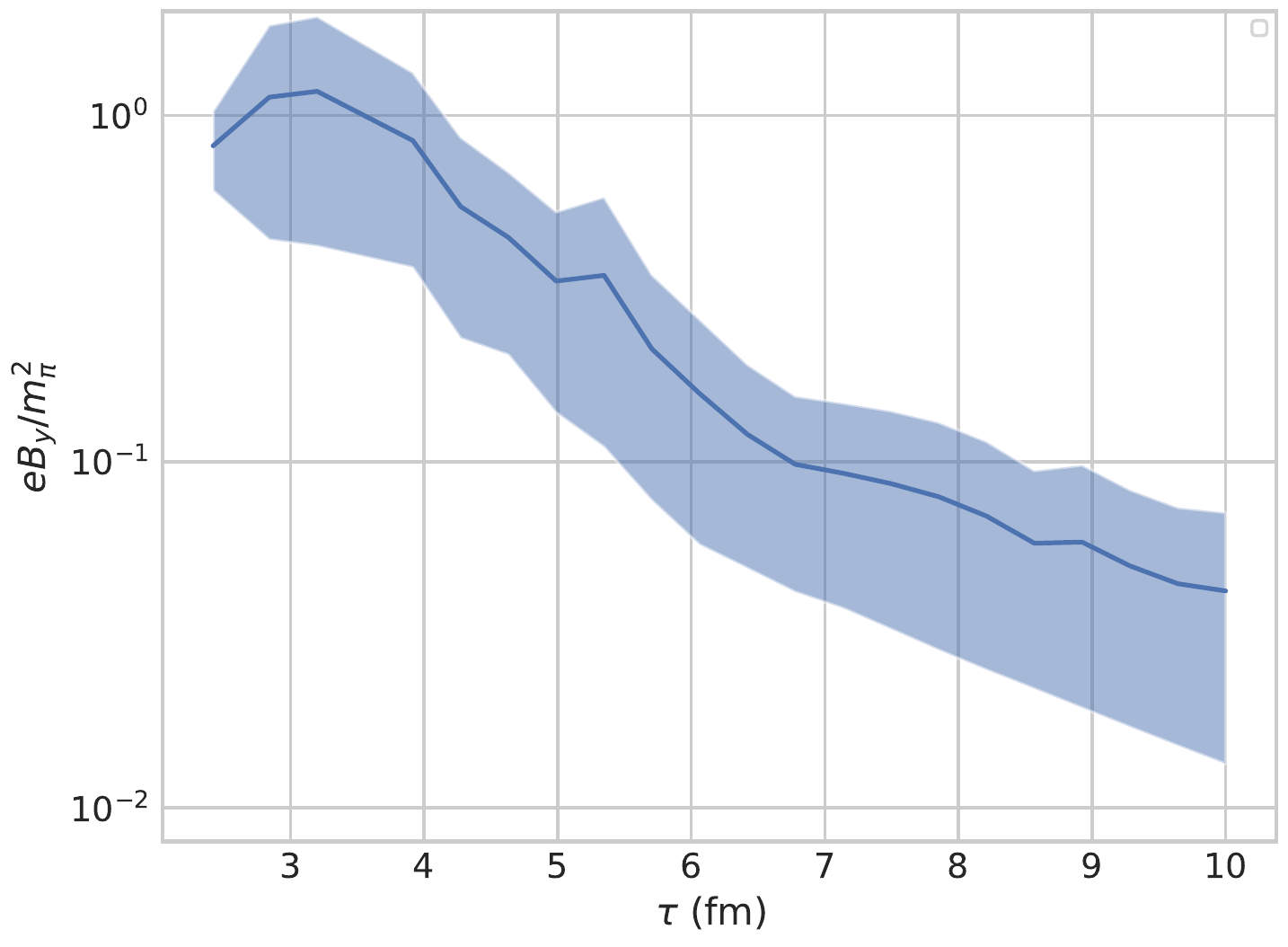} 
		\caption{Time evolution of $eB_y$ at rapidity $\eta=0.5$. The blue band is the estimate of the error in the numerical integration.} \label{fig:by1}
	\end{figure}
	
	Fig.\eqref{fig:Eeta1} shows the variation of $e \tilde{E}_{\eta}$ with respect to $\eta$ at different time frames. Compared to the electric field generated by a point charge distribution Fig.\eqref{Electricfield_taueta} (left panel), the electric field generated by a charge distribution has support even for time as early as  $\tau \sim $ 1.1 fm although still localized in space. At late times, the electric field asymptotically goes to zero at large rapidities. The magnitude of  of electric field is of the order of $\sim 0.1 m_{\pi}^2$ and this acts as the source for magnetic field. Fig.\eqref{fig:by1} shows the temporal evolution of transverse component of magnetic field $eB_y$ at rapidity $\eta=0.5$ and the blue band is an estimate of the error in the numerical integration. Since, the integrand in Eq.\eqref{Eq:Eetasol1} is highly oscillatory, we could not extrapolate to smaller time interval regions, nevertheless the qualitative behavior of magnetic field at late time $\tau_f>4.6$ fm is similar to that of a point charge distribution 
	Fig.\eqref{Magneticfield_taueta} (right panel) and remains non-vanishing for times long enough in the context of heavy-ion collisions.  The long life time is the result of retardation effect which can be readily seen from Fig.\eqref{Domain_taueta} and emphasizes the fact that for late time the support of the integral in Eq.\eqref{Eq:Eetasol1} increases to larger spatial regions which has non-vanishing gradients of electric field. 
	\section{Conclusion and outlook} \label{sec:conclusion}
	In this study, we investigated the spatiotemporal behavior of electromagnetic fields generated by charged particles in an expanding fluid under a background Bjorken flow. We solved Maxwell's equations in this background, assuming no back-reaction to the flow. The inclusion of coupling to the fluid's shear and expansion scalar has made the dynamics much more intricate. An important finding of this study is that even in the absence of charged currents, the gradient of the electric field can produce a magnetic field for a stationary charge distribution co-moving with the fluid. The resulting magnetic field initially vanishes, grows, and eventually decays. Causality plays a decisive role  for describing the space-time evolution for such charge distributions. We also discussed a more realistic case of continuous charge distribution in the context of heavy-ion collision. The resulting magnetic field remains appreciable even for time as large as $\sim10$ fm. This finding supports the works that discuss the effect of magnetic field in the hadronic stage of heavy-ion collision~\cite{Dash:2020vxk,Dey:2020awu,Imaki:2019rlm}.
	
	A major limitation of this study is the assumption of vanishing charged currents. This could have significant and interesting consequences on the results obtained in this study when non-equilibrium processes such as charge diffusion and conductivity come into play. These processes eliminate any gradients present in the electric charge distribution and may have non-trivial effects on the space-time variation of dynamic electromagnetic fields. When such terms were included, we were unable to find a closed analytical solution to the Green's function and are therefore difficult to solve. Other possible directions for future work could include considering flow fields with non-vanishing vorticity and acceleration. These and other intriguing phenomena will be left for future investigations.

	\begin{acknowledgments}
		We thank A. Mukherjee, D. Rischke and V. Roy for fruitful discussions. A.D gratefully acknowledges the support of the
		Alexander Von Humboldt Foundation through a research fellowship for postdoctoral researchers. A.P acknowledges the CSIR-HRDG financial support. 
	\end{acknowledgments}


	\bibliography{ref2}

\end{document}